\documentstyle[12pt]{article}
\newcommand{\la}{\label}

\newcommand{\n}{\vec {\bf n}}
\newcommand{\w}{\vec {\bf w}}

\newcommand{\x}{\vec {\bf x}}

\newcommand{\D}{{\cal D}_\tau}

\newcommand{\be}{\begin{equation}}
\newcommand{\ee}{\end{equation}}
\newcommand{\ba}{\begin{eqnarray}}
\newcommand{\ea}{\end{eqnarray}}
\newcommand{\bastar}{\begin{eqnarray*}}
\newcommand{\eastar}{\end{eqnarray*}}

\begin{document}
\begin{titlepage}
 
 
 
\begin{center}
{ 
\bf \Large \bf KNOTS IN INTERACTION 
}
\end{center}
 
\vskip 2.5cm
 
\begin{center}
{\bf Antti J. Niemi} \\
\vskip 0.3cm

{\it $^*$Department of Theoretical Physics,
Uppsala University \\
P.O. Box 803, S-75108, Uppsala, Sweden } \\
 
\vskip 0.3cm

{\it Helsinki Institute of Physics \\
P.O. Box 9, FIN-00014 University of Helsinki, Finland} \\

\vskip 0.3cm

{\it The Mittag-Leffler Institute  \\
Aurav\"agen 17, S-182 62 Djursholm, Sweden} 
\end{center}

\vskip 2.5cm
\rm
\vskip 0.5cm {\rm {\small 
We study the geometry of interacting knotted solitons.
The interaction is local and advances either
as a three-body or as a four-body process, depending 
on the relative orientation and a degeneracy of the 
solitons involved. The splitting and adjoining is governed
by a four-point vertex in combination with duality 
transformations. The total linking number 
is preserved during the interaction. It receives 
contributions both from the twist and the writhe, 
which are variable. Therefore solitons  
can twine and coil and links can be formed.} }

\noindent\vfill

\begin{flushleft}
\rule{5.1 in}{.007 in} \\
$^{*}${ \small permanent address }\\ \vskip 0.2cm
{\small Supported by NFR Grant F-AA/FU 06821-308}
\\ \vskip 0.3cm
\hskip 0.0cm {\small  E-mail: \scriptsize
\bf NIEMI@TEORFYS.UU.SE}  \\
\end{flushleft}
\end{titlepage}

\vskip 1.0cm

Knotlike configurations appear in 
a variety of physical, chemical and biological 
scenarios. Examples include Early Universe 
cosmology, the structure of elementary 
particles, magnetic materials, turbulent 
fluid dynamics, polymer folding, and 
DNA replication, transcription 
and recombination. Whenever knots
occur, it becomes important 
to understand how they twist, coil, split and 
adjoin. 

Here we outline a geometrical first principles
approach for understanding knots in interaction.
For definiteness we shall consider Hamiltonian 
field theories where knots
appear as solitons {\it i.e.}
as stable finite energy solutions to the pertinent
nonlinear equations of motion \cite{nat}-\cite{hie}. 
Even though our 
conclusions are independent of any detailed 
stucture of the equations of motion,
for concreteness we specify to a three component vector 
field $\vec{\bf n}(\x)$ with unit 
length $ \vec{\bf n} \cdot \vec{\bf n} 
= 1$ \cite{fad}. We note that $\n$ appears as an order parameter in 
a variety of applications. It can for example describe 
magnetization in a ferromagnet, velocity distribution 
$\n = \vec {\bf v}/| \vec {\bf v}| $ 
in fluids, the director field in
a nematic liquid crystal, and quantum fluctuations 
of a Higgs field around a spontaneously 
broken vacuum in a grand unified field 
theory. 

A soliton configuration is localized, and
$\n$ goes to a constant vector $\vec{ \bf n}(
\vec {\bf x}) \to \vec{\bf n}_0$ at spatial infinity. 
Therefore it is a map from the 
compactified three-space $R^3 \sim S^3$ 
to the internal unit two-sphere $S_{\n}^2$. 
Such maps fall into nontrivial homotopy 
classes $\pi_3(S^2) \simeq Z$ 
labelled by the Hopf invariant \cite{rol} which 
can be evaluated by noting 
that the pre-image in $R^3$ of a point in $S_{\n}^2$
is an embedding of a circle $S^1 \in R^3$. This 
embedding is a knot in the mathematical 
sense \cite{rol}, and the  
Hopf invariant is the linking number of any 
two such knots. We select the asymptotic vector $\vec{\bf 
n}_0$ to point down, along the negative $z$-axis 
in three-space. At the core (center) of 
the knotted soliton $\n$ then points up, and the pre-image
of the core is a mathematical knot 
$\vec {\bf x}(\sigma)$ parametrized 
by $\sigma \in [0, 2\pi]$ so that
$\vec {\bf x}(\sigma + 2\pi) = \vec {\bf x}(\sigma)$. 

We consider a short strand of a 
knotted soliton, a tubular distribution of 
energy density that surrounds 
the core and does not overlap with any other strand. 
We inspect it with a planar cross 
sectional disk $\D \in R^3$ 
that cuts the strand at right angle to 
its core at $\x(\sigma)$. The center of $\D$ coincides 
with the core, and its boundary is a circle 
around the core where $\n \to \n_0 = 
-{\bf \hat{\bf e}}_{z}$, see figure 1. 
Note that topologically each $\D$ is a 
(Riemann) two-sphere $S^2_{\sigma}$, and
$\n$ has the structure of
a map $S^1 \times S^2_{\sigma} \to S^2_{\n}$ that 
can be locally parametrized by
\be
\n(\x) \ = \ 
\left( \begin{array}{c} 
n_1 \\
n_2 \\ 
n_3 
\end{array} \right) 
\ = \
\left( \begin{array}{c} 
\cos( k \tau + l \varphi) \sin \theta \\
\sin( k\tau + l \varphi) \sin \theta \\ 
\cos \theta
\end{array} \right) 
\la{n}
\ee
Here $\tau$ is a modular coordinate for the
self-linking along the core, it 
is proportional to $\sigma$. The 
integer $k$ measures the net number of twists 
in our strand when we proceed along the core.  
The coordinates $\varphi$ and $\theta$ 
are defined at each $\sigma$, and cover 
the disk $\D$. The angle $\varphi \in 
[0,2\pi]$ is the azimuth around the 
center of $\D$ and $\theta \in 
[0,\pi]$ is the polar 
angle with $\theta = 0$ at the core 
and $\theta = \pi$ at the boundary of $\D$. 
The integer $l$ labels the $\pi_2(S^2) 
\simeq Z$ homotopy class of $\n$ as a map
from the disk $\D \sim S^2_{\tau}$ 
to the internal two-sphere $S^2_{\n}$.

Our interpretations become more apparent 
when we substitute (\ref{n})
in the integral representation of the 
Hopf invariant \cite{nat},
\be
Q_H \ = \  \frac{k \cdot l}{8\pi^2} \int 
\sin \theta d\theta \wedge d\varphi \wedge d\tau 
\la{hopf}
\ee
Notice that a knotted soliton 
has either a right-handed or a left-handed 
orientation which is determined 
by $d\theta \wedge d\varphi \wedge d\tau$. 
Furthermore, for {\it any} regular 
Hamiltonian the Hopf invariant is conserved {\it even} 
when we have interactions where strands split and
adjoin. 

We combine $n_1$ and $n_2$ in (\ref{n}) into a 
two component vector field $\w = (n_1 , n_2)$. 
The core $\theta = 0$ is a fixed point of $\w$. 
Therefore we can define an 
index $Ind_{\varphi}(\w)$ that counts how many 
times $\w$ rotates around the core when we 
go once around it along some curve on 
the disk $\D$ in the right-handed direction. This
index is a topological invariant which is 
independent both of the curve on $\D$ and of 
the point $\sigma$ along the core, 
and from (\ref{n}) we find
$Ind_{\varphi}(\w) = l$. This index is 
also additive, if we evaluate 
it along a curve that surrounds $N$ fixed 
points of $\w$ the result coincides with the 
sum of the $N$ indices for the individual 
fixed points.
 
The third component $n_3$ in (\ref{n}) is a 
continuous function on $\D$ with values 
between $[-1,+1]$. It has a maximum 
at the core $\theta = 0$ and 
its minimum occurs on the boundary of the disk 
$\D$ where $\theta \to \pi$. 
In general the maximum at $\theta = 0$ 
is a degenerate critical point, and for 
our knotted solitons we expect that its 
degeneracy has a definite multiplicity 
given by the absolute value of the index 
$Ind_{\varphi}(\w)$,  
\be
deg(n_3)_{|\theta = 0} \ = \ \left|
Ind_{\varphi}(\w)\right| \ \sim  \ |l|
\la{deg}
\ee
In particular, when $Ind_{\varphi}(\w) =  
\pm 1$ the core is non-degenerate. 
In the model studied in \cite{nat}-\cite{hie} we 
can verify (\ref{deg}) explicitly: 
If $\rho$ measures distance from the core 
in local polar coordinates $(\vartheta,\rho)$ on $\D$, 
we find from the equations of motion that 
the core is truly a degenerate critical point
of $n_3$ on $\D$ with $l$-fold multiplicity,
\be
n_3 (\vartheta,\rho) \ \stackrel{\rho \to 0}{=}  
\ 1 - {\cal O}(\rho^{2|l|})
\la{est1}
\ee
This result is related to analyticity 
near $\theta = 0$: The disks $\D$ 
have a natural complex structure 
$\sin \theta d\theta \wedge d \varphi
\sim dz \wedge d{\bar z}$ suggesting
that (\ref{deg}) should be valid whenever 
the equations of motion are elliptic.  
Indeed, in our physical applications
(\ref{deg}) makes sense.  
It means that a knotted soliton with $l$-fold 
degeneracy is a bound state of $l$ 
non-degenerate ones.

Since $\theta = 0$ is a critical point of $n_3$,
we have
\be
{\partial n_3 \over 
\partial x_i } (\theta = 0) \ = \ 0
\la{inte1}
\ee
and since the core is a curve 
in $R^3$, the $3 \times 3$ symmetric matrix 
\be
{\partial^2 n_3 \over \partial x_i \partial x_j }(
\theta = 0) 
\la{inte2}
\ee
has one vanishing eigenvalue $\lambda_1 = 0$,
the corresponding eigenvector is tangent to the core. 
Since $\theta = 0$ is a maximum, the two 
additional eigenvalues of (\ref{inte2}) are 
non-positive $\lambda_{2,3}(\tau) \leq 0$ at the core.
If both are non-vanishing the core is a 
non-degenerate critical point of $n_3$ on $\D$. 
But if the core is a degenerate
critical point then either $\lambda_2$ 
or $\lambda_3$ (or both) should vanish.  

We now proceed to apply these  
considerations to geometrically 
describe the interactions 
between strands of knotted solitons. We first 
inspect the situation with two 
initial strands that both have a 
non-degenerate core. We then explain how
the results generalize when
strands have degenerate cores.

When both initial strands have a non-degenerate
core, the corresponding eigenvalues $\lambda_{2,
3}(\tau)$ are non-vanishing. Therefore a nontrivial 
interaction such as splitting and adjoining can not 
occur unless at least one of the $\lambda_{2,3}(\tau)$ 
vanishes. At that point the two cores must coincide. 

We first assume that the two initial strands 
have a parallel relative orientation. 
From (\ref{deg}) we then conclude that when  
the cores coincide $\theta = 0$ becomes 
a doubly degenerate critical point of $n_3$
with $Ind_{\varphi}(\w) = \pm 2$, depending 
on whether the two strands are right-handed or 
left-handed. The interaction 
is pointwise with two outgoing 
non-degenerate strands, and it can be described by 
a {\it four-point} vertex. 

To visualize the vertex we 
draw plane projections of the strands.
For this we trace the three gradient vectors 
$d\theta$, $d\varphi$ and $d\tau$ 
along the cores. The gradient $d\tau$ 
is (co)tangent to a core. In a plane projection 
we describe it by drawing an arrow along the direction 
of the core. For each $\tau$ the two additional 
gradients $d\theta$ and $d\varphi$ span the cross 
sectional planes $\D$. We describe them by drawing 
two oriented lines in the vicinity of the core.

We recall that the self-linking number $Lk$ 
of a knotted soliton can be computed 
from its plane projection using \cite{rol}
\be
Lk  \ = \ Tw \ + \ Wr
\la{LTW}
\ee
where $Tw$ is the twist and $Wr$ 
is the writhe in the plane projection.
During an interaction neither 
$Tw$ nor $Wr$ is in general separately conserved. 
Only their sum is since it coincides with the
Hopf invariant which is conserved.

To picture the interaction vertex, we first 
employ continuity to translate all twists and 
writhes in the initial strands away from the vicinity 
of the interaction point. The plane projections 
of the initial configurations then become 
well-groomed lines with either antiparallel (A) or 
parallel (P) alignment. For example, in the case of two 
(right-handed) planar strands with
antiparallel (A) alignment, we have the four-point 
interaction vertex $V_A$ in figure 2. Here both 
the twist and the writhe are separately conserved. 
In the case of two (right-handed) planar strands 
with parallel (P) alignment we have the four-point 
interaction vertex $V_P$ in figure 3. Here neither 
the twist nor the writhe are conserved but
their sum is, since the interaction preserves 
the Hopf invariant. Note that a difference in 
the final twist and writhe between the two vertices
has important physical consequences, it suggests that
the strands will twine, coil and link. Indeed, by 
repeating $V_P$ twice as in figure 4, we 
either continue coiling or form a link, depending on
the global geometry. This is not equally obvious in the case
$V_A$. However, we note that in $R^3$ 
a writhe can always be continuously deformed into a twist. 
Furthermore, two non-coplanar strands can exhibit an 
antiparallel alignment in one plane projection, 
but parallel alignment in another plane projection. 
Consequently we expect that the two vertices $V_P$ and $V_A$
should be related. For this we first vertically flip 
the plane projection of one of the two initial 
strands in figure 3. This yields two strands 
with antiparallel alignment in the plane projection. 
We then implement an interaction described by the 
vertex $V_A$ in figure 2, followed by another vertical 
flip. The result coincides with that obtained by implementing
the vertex $V_P$ alone. This means that  
the two vertices $V_A$ and $V_P$ are dual to each 
other, with vertical flip defining the duality 
transformation. 
 
Depending on the dynamical details of the Hamiltonian, 
besides splitting and adjoining the two initial 
strands may also combine into a 
single strand but with a doubly degenerate core. This
can be described using the topologically invariant 
index $Ind_{\varphi}(\w)$. As an example we consider 
two initial right-handed strands both with 
$Ind_{\varphi}(\w) = +1$ (figure 5). 
Since the index is additive, we conclude that for a curve 
which encircles both cores we have $Ind_{\varphi}(\w) 
= +2$. This is also the index of the final, doubly 
degenerate strand. The interaction is a process 
where two homotopically nontrivial maps $S^1_{\varphi} 
\to S^1$ each with a $\pi_1(S^1)$ winding number $+1$ 
combine into a single map with $\pi_1(S^1)$ winding 
number $+2$. Dynamically, this 
leads to the formation of a doubly degenerate 
knotted soliton, a bound state,  in a manner  
which is consistent with the conservation of the 
Hopf invariant. We visualize the interaction  
as a three-body process in figure 5.

\vskip 0.3cm
When the two initial strands have an 
opposite orientation, splitting and adjoining 
can not occur since two strands with different 
orientation should not connect. Instead 
the two strands can annihilate each other.
As in figure 5, this can also be described 
using the topologically invariant 
index $Ind_{\varphi}(\w)$. For example, if 
one of the initial strands corresponds to 
$Ind_{\varphi}(\w) = +1$ and the other 
strand to $Ind_{\varphi}(\w) = -1$, since the index is 
additive we conclude that for a curve that 
encircles both cores we have $Ind_{\varphi}(\w) =0$. 
As in figure 5, the interaction is a 
process where two homotopically nontrivial 
maps $S^1_{\varphi} \to S^1$ combine. 
But now one of the initial maps has $\pi_1(S^1)$ 
winding number $+1$ and the other $-1$, and the
combined map is homotopically trivial.
This corresponds to an annihilation between the 
two strands, that proceeds dynamically in a manner
which is consistent with the conservation of the 
Hopf invariant. The annihilation can be 
visualized as a three-body process, much like in
figure 5. 

In the general case where the initial strands 
are degenerate the interaction can proceed 
in a multitude of different fashions, details 
depending on the degrees of degeneracy and 
the relative orientations of the 
interacting strands. But according 
to (\ref{deg}) degenerate strands can  
always be viewed as bound states of non-degenerate 
strands. Thus there is no need to discuss their
interactions separately. We can always describe 
their interactions using various combinations 
of the interactions that occur between non-degenerate 
strands, in a rather obvious manner.

Finally, the interaction processes that we have 
outlined here all conserve the Hopf invariant. 
This follows from the condition that the order 
parameter $\n$ is a unit vector. But
in a number of applications one 
may wish to relax the constraint $\n \cdot \n =1$ 
so that $|\n|$ can vanish at some points in $R^3$.
One way to achieve this is to replace the constraint 
by a term $\lambda (\n \cdot \n -1)^2$ in the 
Hamiltonian, with $\lambda$ a coupling constant. 
When $\lambda \to \infty$ we recover $\n \cdot 
\n =1$. But for finite (large) coupling we expect
$|\n|$ to fluctuate in the vicinity of $|\n| \approx 1$
even though we can also have $|\n(\x)|\to 0$ at some points.  
This generalizes the models discussed here in 
a manner that should be 
of interest in a variety of applications. For example 
in quantum field theory it allows us to recover
renormalizability, while in molecular biology it
enables us to account for effects of entzymes in DNA. 
Obviously study of interaction processes
with a finite $\lambda$ should be of 
wide interest.

\vskip 0.4cm
We have found that depending
on the relative orientation and degeneracy of
the solitons involved, an actual interaction can 
involve various combinations of splitting 
and adjoining, twisting, coiling and linking. 
It can lead to the formation 
of bound states and annihilation between 
strands. Our description of the interaction processes
is quite model independent and based entirely
on elementary conceps of continuity and differentiability.
Consequently we hope that our results can form a 
basis for detailed investigations of interacting knots.
Indeed, we expect that numerical studies 
in the model discussed in  \cite{nat}-\cite{fad}
(and its generalization to the case where $|\n|$ can vanish)
provide prolific tests for various 
aspects of knot interaction.

\vskip 1.0cm
I thank Ludvig Faddeev for many discussions
on knots and solitons. I am particularly indebted
to Oleg Viro for several very helpful conversations 
and his guidance that prompted the present work. 
I also thank A. Alekseev, R. Battye, J. Hietarinta, 
L. Kauffmann, N. Nekrasov, A. Polychronakos, 
A. Schwarz and P. Sutcliffe for discussions. 
This work has been supported by NFR Grant 
F-AA/FU 06821-308

\vfill\eject

\vskip 1.0cm
\begin{flushleft}
{\bf Figure Caption}
\end{flushleft}
\vskip 0.5cm

{\bf Figure 1:} A cross-sectional disk $\D$ cuts a strand
at a right angle to its core. The center of the disk 
coincides with the core where $\n$ points up, and its
boundary is a circle where $\n$ points down. 

\vskip 0.4cm
{\bf Figure 2:} The plane projected interaction 
vertex $V_A$ between two right-handed, mutually 
antiparallel strands.

\vskip 0.4cm
{\bf Figure 3:} The plane projected interaction vertex
$V_P$ between two right-handed, mutually parallel strands. 

\vskip 0.4cm
{\bf Figure 4:} If repeated twice, the vertex $V_P$ can
lead to further coiling or to the formation of a link.

\vskip 0.4cm
{\bf Figure 5:} Two right-handed non-degenerate strands 
can combine into a single doubly-degenerate strand.
The process leads to a doubly degenerate knotted soliton.


\begin{thebibliography}{9}

\bibitem{nat} L. Faddeev and A.J. Niemi, {\it Nature} {\bf 387}, 
58 (1997) 

\bibitem{gla} J. Gladikowski and M. Hellmund, {\it 
Physical Review} {\bf D56} 5194 (1997)

\bibitem{sut} R. Battye and P. Sutcliffe, hep-th/9811077,
hep-th/9808129

\bibitem{hie} J. Hietarinta and P. Salo, hep-th/9811053

\bibitem{fad} L. Faddeev, {\it Quantisation of Solitons},
preprint IAS Print-75-QS70 ,1975; and in  {\it Einstein
and Several Contemporary Tendencies in the Field Theory
of Elementary Particles} in Relativity, Quanta and 
Cosmology vol. 1, Pantaleo M. and De Finis F. (eds.),
Johnson Reprint, 1979

\bibitem{rol} D. Rolfsen, {\it Knots and Links} (Publish 
or Perish, Berkeley CA 1976); L.H. Kauffman, {\it Knots And Physics} (World 
Scientific, Singapore 1993)

\end{thebibliography}
\end{document}